\documentclass[aps,prd,10pt,showpacs,amsmath,showkeys,twocolumn,floatfix,amssymb, preprintnumbers, nofootinbib, superscriptaddress,longbibliography]{revtex4-1}
\usepackage{graphicx}
\usepackage{comment}
\usepackage[usenames]{color}
\usepackage{bm}
\usepackage{ifpdf}
\usepackage{floatrow}
\usepackage{makecell}
\usepackage{caption}
\usepackage{stackrel}
\usepackage{subcaption}
\usepackage{enumitem}

\usepackage[normalem]{ulem}
\usepackage[dvipsnames]{xcolor}
\usepackage[utf8]{inputenc}
\usepackage{hyperref}
\hypersetup{
  pdfnewwindow=true,      
  colorlinks=true,        
  linkcolor=PineGreen,    
  citecolor=PineGreen,    
  filecolor=PineGreen,    
  urlcolor=PineGreen      
}
\newcommand{\be}{\begin{eqnarray}}
\newcommand{\ee}{\end{eqnarray}}

\begin{document}

\title{Toward extracting scattering phase shift from integrated correlation functions on quantum computers}

\author{Peng~Guo}
\email{peng.guo@dsu.edu}

\affiliation{College of Arts and Sciences,  Dakota State University, Madison, SD 57042, USA}
\affiliation{Kavli Institute for Theoretical Physics, University of California, Santa Barbara, CA 93106, USA}

\date{\today}

\begin{abstract}
Based on a established relation   that relates  the integrated correlation functions for a trapped system to the infinite volume scattering phase shifts through a weighted integral,  we propose to extract  the infinite volume scattering phase shifts through quantum simulation of the integrated correlation functions of trapped two-particle systems on quantum computers.    The integrated correlation function   can be computed by an ancilla-based algorithm. The proposal is demonstrated with a simple contact interaction fermion model.  
\end{abstract}

\maketitle

\paragraph*{Introduction.\textemdash}
Few-body interaction and scattering processes play a key role in a large variety of physics phenomena ranging from nuclear reactions in nuclear and astrophysics  to electron-phonon scattering  in material sciences.   The determination of  few quantum particles interaction and scattering amplitudes is a fundamental problem  in many areas of modern physics,  including nuclear, atomic and condensed matter physics.    Numerical simulations of scattering processes  of strongly coupled quantum systems, such as strong interaction between quarks and gluons in nuclear physics  and strongly correlated electrons system in condensed matter physics,  remain   challenging problems.  Classical Monte Carlo simulations suffer a number of fundamental issues, including  lack of access to real-time dynamics, critical slowing down \cite{SCHAEFER201193}, infamous sign problems  in fermion systems \cite{PhysRevB.41.9301,deForcrand:2009zkb}, and signal-to-noise ratio issues \cite{lepage1989analysis,DRISCHLER2021103888} in lattice  QCD calculations of few-nucleon systems.   Despite  challenges, a lot of progresses have been made in recent years on determining and extracting few-hadron interactions in nuclear and hadron physics.  Typically,  computations based on stochastic evaluation of the path integral  are   performed in an artificial trap,  such as a harmonic oscillator trap in nuclear physics or  a periodic finite box in lattice quantum chromodynamics (QCD) calculation,  the discrete low-lying energy spectra are extracted by evaluating exponential decay of few-body correlation functions in Euclidean space-time.     Eigenenergies are  then converted into scattering phase shifts by applying L\"uscher formula  \cite{Luscher:1990ux} in periodic boxes   or  Busch-Englert-Rza\.zewski-Wilkens (BERW) formula \cite{Busch98}  in harmonic oscillator traps, also see extension of such formula in   inelastic  and three-body channels,  e.g. Refs.~\cite{Rummukainen:1995vs,Christ:2005gi,Bernard:2008ax,He:2005ey,Lage:2009zv,Doring:2011vk,Guo:2012hv,Guo:2013vsa,Kreuzer:2008bi,Polejaeva:2012ut,Hansen:2014eka,Mai:2017bge,Mai:2018djl,Doring:2018xxx,Guo:2016fgl,Guo:2017ism,Guo:2017crd,Guo:2018xbv,Mai:2019fba,Guo:2018ibd,Guo:2019hih,Guo:2019ogp,Guo:2020wbl,Guo:2020kph,Guo:2020iep,Guo:2020ikh,Guo:2020spn,Guo:2021lhz,Guo:2021uig,Guo:2021qfu,Guo:2021hrf,Stetcu:2007ms, Stetcu:2010xq,Rotureau:2010uz,Rotureau:2011vf,Luu:2010hw,Yang:2016brl,Johnson:2019sps,Zhang:2019cai, Zhang:2020rhz}.

 Quantum computers have emerged as a promising alternative to simulate quantum systems more efficiently and offer the prospect   to overcome   some of the limitations in classical simulations.  Number of quantum algorithms  toward computing two-body scattering on quantum computers have been proposed recently, such as (1)  the   L\"uscher formula-like  approach  in Ref.~\cite{PhysRevC.109.064623,PhysRevC.110.054604}:  extracting   eigenenergies  by Variational Quantum Eigensolver in quantum computing  and then   converting them into scattering phase shifts by L\"uscher-  or BERW-like formula;  (2) measuring phase shift by wave packet time delay \cite{PhysRevD.104.054507}; and (3) reconstructing scattering amplitudes through the coupling of the particle with an ancillary spin-1/2 \cite{Mussardo2024}.

In present work, we propose  the direct computation of the integrated correlation functions of trapped particles systems on quantum computers,    then  the scattering phase shifts in infinite volume can be established  by the difference of integrated correlation functions between interacting and non-interacting particles systems in the trap through a weighted integral \cite{Guo:2023ecc,Guo:2024zal,Guo:2024pvt}.  This letter is organized as follows:  (i) we firstly summarize the formalism that relates the integrated correlation functions for  a particles system in a trap to the infinite volume scattering phase shifts; (ii) next we show how the Hamiltonian of trapped system and correlation functions can be mapped onto quantum circuits that can be evaluated on quantum computers; (iii) finally the feasibility  of the proposed approach is demonstrated with a simple contact interaction fermion model  followed by the discussion and summary.

\paragraph*{Two-particle scattering and integrated correlation function formalism.\textemdash}
    
In this section, we briefly outline and summarize the fundamental of integrated correlation function formalism that offers an alternative approach for extracting the infinite volume scattering phase shifts from few-body systems in a trap. The complete technical details can be found in Refs.~\cite{Guo:2023ecc,Guo:2024zal,Guo:2024pvt}.

A simple  $1+1$  dimensional  nonrelativistic field theory model  of spin-1/2 fermions interaction via a contact interaction in a trap is  used in this work,    the Hamiltonian operator of the trapped fermions system is  
\begin{align}
\hat{H} &=\sum_{\sigma = \uparrow, \downarrow } \int d x \hat{\psi}^\dag_{\sigma} (x) \left [ - \frac{1}{2m} \frac{d^2}{d x^2}  + U(x)\right ] \hat{\psi}_{\sigma} (x)  \nonumber \\
&  + V_0 \int d x    \hat{\psi}^\dag_{\uparrow} (x) \hat{\psi}^\dag_{\downarrow} (x)    \hat{\psi}_{\downarrow} (x) \hat{\psi}_{\uparrow} (x),
\end{align}
where $\sigma= \uparrow, \downarrow $ and $m$ refer to the fermion polarizations and mass respectively, and $ \hat{\psi}_{\sigma} (x) $ stands for the fermion field operator. $U(x)$ represent the trap potential, and  $V_0$ is the strength of contact interaction potential  between two fermions with opposite polarizations.  The commonly used artificial traps include periodic finite box in lattice QCD and harmonic oscillator trap in nuclear physics community.  

As illustrated in Refs.~\cite{Guo:2023ecc,Guo:2024zal,Guo:2024pvt},  the infinite volume two-particle elastic scattering phase shift, $\delta(\epsilon)$, can be related to the integrated correlation functions for two-particle in a trap through a weighted integral
 \begin{equation}
 C(t) - C_0 (t) \stackrel{\text{trap} \rightarrow \infty}{\rightarrow} \frac{i\,t}{\pi} \int_0^{\infty} d \epsilon  \,\delta(\epsilon)\, e^{ -i\,\epsilon\, t}, 
 \label{mainresult}
 \end{equation} 
 where $C(t)$ and $C_0 (t)$ are integrated correlation functions for two interacting and non-interacting particles in the trap.   The difference of integrated  trapped two-particle correlation functions converge rapidly  to its infinite volume limit that is given by  an weighted integral over scattering phase shift   at short time.

 The   integrated forward time propagating two-particle correlation function for non-relativistic systems is defined through summing over all the modes of  two-particle correlation functions  along the diagonal,
\begin{equation}
C(t) = \int d r\,     \langle 0 |   \widehat{\mathcal{O}} (r) e^{- i \hat{H} t}  \widehat{\mathcal{O}}^\dag (r)   | 0 \rangle. \label{Ctdef}
\end{equation}
The  $\widehat{\mathcal{O}}^\dag (r)$ and $\widehat{\mathcal{O}} (r)$ denote creation  and annihilation operators  to create two particles with relative coordinate of $r$ at time $0$ at source, and then annihilate them with relative coordinate of $r$ at later time $t$ at sink, respectively.    For instance,    the total spin-singlet   two spin-$\frac{1}{2}$ ferminons  operator after projecting out center of mass motion in a periodic box trap is given by
\begin{align}
& \widehat{\mathcal{O}}^\dag (r) \nonumber \\
& = \frac{1}{\sqrt{ L}} \int_0^L d x_2 \frac{ \psi^\dag_\uparrow (r + x_2) \psi^\dag_\downarrow (x_2 )- \psi^\dag_\downarrow (r + x_2) \psi^\dag_\uparrow (x_2 ) }{2}, \label{twobodyoperator}
\end{align}
where  $L$ stands for the length of a periodic box, and  $ \hat{\psi}_{\sigma} (x) $ is required to satisfies periodic boundary condition: $ \hat{\psi}_{\sigma} (x + L) = \hat{\psi}_{\sigma} (x) $.

We have also shown in Refs.~\cite{Guo:2023ecc,Guo:2024zal,Guo:2024pvt} that the energy spectral representation of  integrated correlation function is
 \begin{equation}
C(t)   = Tr \left [ e^{- i \hat{H} t} \right ] =  \sum_\epsilon  e^{-  i\,\epsilon \, t }  , \label{Cenergyspectrarep}
\end{equation}
where the energy spectrum of trapped interacting systems, $\epsilon $, can be determined by L\"uscher formula-like quantization conditions.

\paragraph*{Second quantization representation of trapped system and fermion-to-qubit mapping.\textemdash}

One of the major advantages of putting particles in a trap is that      the  Hamiltonian operator of interacting particles systems  can be written in the trap basis with discrete energy spectra,  which is   convenient and suitable  for the further mapping onto quantum gate operations in quantum computing.

By expanding field operators in  the trap basis,
\begin{equation}
\hat{\psi}_{\sigma} (x) = \sum_n \varphi_n (x) a_{n, \sigma}, \ \ \ \hat{\psi}^\dag_{\sigma} (x) = \sum_n \varphi^*_n (x) a^\dag_{n, \sigma},
\end{equation}
where $a_{n, \sigma}$ and $a^\dag_{n, \sigma}$ are the annihilation and creation operators for a non-interacting  fermion state   with eigen-energy of $ \epsilon_n^{(0)}$ in a trap, and $ \varphi_n (x) $ are the eigen-solutions of Hartree-Fock-like equation,
\begin{equation}
\left [ - \frac{1}{2m}  \frac{d^2}{dx^2} + U(x) \right ]  \varphi_n (x)  = \epsilon_n^{(0)}  \varphi_n (x) ,
\end{equation}
  the   second quantization representation of Hamiltonian of trapped fermions system  is thus given by
\begin{align}
\hat{H} & =\sum_{\sigma  , n }   \epsilon^{(0)}_n  a^\dag_{\sigma, n}  a_{\sigma, n}   \nonumber \\
& +    \sum_{n_1, n_2, n'_1 , n'_2} V_{n'_1, n'_2; n_1, n_2}  a^\dag_{\uparrow, n_1}  a^\dag_{\downarrow, n_2}  a_{\downarrow, n'_2 }  a_{\uparrow, n'_1} , \label{Hamiltonian}
\end{align}
 where
 \begin{equation}
 V_{n'_1, n'_2; n_1, n_2}   = V_0  \int d x   \varphi^*_{n'_1} (x) \varphi^*_{n'_2} (x)     \varphi_{n_1}  (x)   \varphi_{n_2} (x)   .
\end{equation}
  The second quantization representation of  Hamiltonian operator  as in Eq.(\ref{Hamiltonian}) can then be mapped onto  qubit quantum registers by applying the Jordan-Wigner transformation, see e.g.  Refs.~\cite{PhysRevA.64.022319,PhysRevA.65.042323,JordanWigner}.

Similarly the two-particle creation and annihilation operators can be expanded in terms of  trap basis and then subsequently mapped onto   qubit quantum registers by applying the Jordan-Wigner transformation. For instance, with a periodic box trap,  the  two-particle creation operator in Eq.(\ref{twobodyoperator})  now has the form:
\begin{align}
 \widehat{\mathcal{O}}^\dag (r) =  \frac{1}{\sqrt{ L}}   \sum_{k    = \frac{2\pi n}{L}, n \in \mathbb{Z}}  e^{- i k r}  \frac{  a^\dag_{k, \uparrow}  a^\dag_{ - k, \downarrow}  - a^\dag_{k, \downarrow}  a^\dag_{ -k,  \uparrow }   }{2} .
\end{align}

\begin{figure}[h]
 \centering
\includegraphics[width=0.95\textwidth]{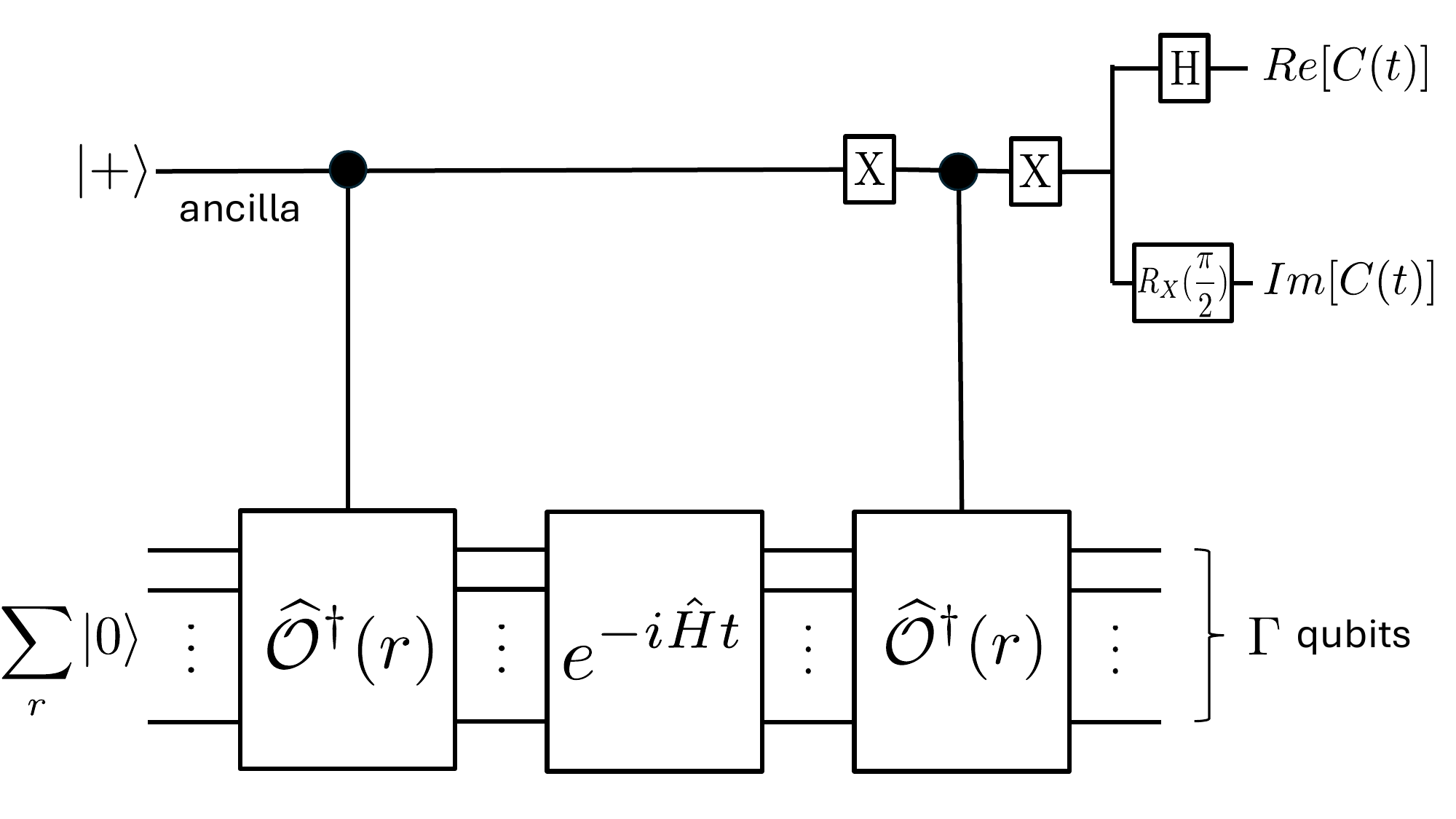}
\captionsetup{singlelinecheck=false, justification=raggedright}
 \caption{Demo   of quantum circuit for computing the integrated two-particle correlation function $C(t)$.  }
 \label{ctplot} 
 \end{figure}

The integrated correlation functions can be computed on quantum computers by an ancilla-based algorithm proposed   in Ref.~\cite{PhysRevA.64.022319}, also see a nice review article in Ref.~\cite{https://doi.org/10.1002/qute.201900052}. The demonstration of quantum circuit for computing  integrated two interacting particles correlation function $C(t)$ is given in Fig.~\ref{ctplot}. The quantum circuit for computing  integrated two non-interacting particles correlation function $C_0(t)$ can be constructed in a similar way by turning the interaction off. The scattering phase shifts then can be extracted from simulation results of $C(t)-C_0 (t)$ by using relation in Eq.(\ref{mainresult}).

\paragraph*{Proof of principles.\textemdash}  
In this section, the proposed approach is demonstrated with fermions interaction in  a periodic box   of size $L$ as a specific example.   Instead of mapping the trapped system onto quantum registers by Jordan-Wigner transformation,  we will show that the mapping and formalism can be simplified drastically by using eigenvector basis of trapped system. With a contact interaction, two-fermion scattering is exactly solvable,  the two-fermion correlation function is given in a simple form that can be checked numerically rather straightforwardly.

\subsubsection{Quantum circuit in trap basis representation}

For a periodic box trap,  the  trap wave function  that satisfies periodic boundary condition   
 is simply given by plane waves: $ \varphi_{k} (x) = \frac{1}{\sqrt{L}} e^{i  k x}$,  where   $k = \frac{2\pi n}{L}, n \in \mathbb{Z}$ are discrete  particle momenta.     Two-particle state in center of mass frame  thus can be defined by
\begin{equation}
| k \rangle =  \frac{  a^\dag_{k, \uparrow}  a^\dag_{ - k, \downarrow}  - a^\dag_{k, \downarrow}  a^\dag_{ -k,  \uparrow }   }{2}  | 0 \rangle , \ \ k = \frac{2\pi n}{L}, n \in \mathbb{Z},
\end{equation}
and the Hamiltonian matrix in $| k \rangle$ basis is  therefore given by a simple form,
\begin{equation}
\hat{H}  =  \sum_{ k  } 2\epsilon^{(0)}_k | k \rangle  \langle k |   + \frac{V_0}{L}  \sum_{ k, k'} | k' \rangle  \langle k | , \label{Hmatkrep}
\end{equation}
where $\epsilon^{(0)}_k = \frac{k^2}{2 m}$.
The integrated correlation function in Eq.(\ref{Ctdef}) now is reduced  to
\begin{equation}
C(t) =  \sum_{k}  \langle k |   e^{- i \hat{H} t}    | k \rangle. \label{ctexpH}
\end{equation}
The Hamiltonian matrix $\hat{H}$ can be decomposed by a diagonalizable matrix $\hat{E}$ and a unitary matrix $\hat{S}$:
\begin{equation}
\hat{H} = \hat{S} \hat{E} \hat{S}^\dag,
\end{equation}
 and   using identity $e^{- i \hat{H} t} = \hat{S} e^{- i  \hat{E} } \hat{S}^\dag$,  the Eq.(\ref{Cenergyspectrarep}) is thus recovered.

 The infinite volume scattering phase shift for a contact interaction is given by, see e.g. Refs.~\cite{Guo:2023ecc}, 
\begin{equation}
\delta(\epsilon)  = \cot^{-1} \left ( - \frac{\sqrt{2\mu \epsilon}}{\mu V_0} \right ), \label{deltaphase}
\end{equation}
where  $\mu = \frac{m}{2}$ is reduced  mass of two fermions system.
Hence,   the analytic expression of right-hand side of Eq.(\ref{mainresult}) can be obtained,
\begin{equation}
C (t) - C_0 (t)    \stackrel{L \rightarrow \infty}{\rightarrow}   \frac{1}{2}  \mbox{erfc} (\mu V_0 \sqrt{\frac{it}{2\mu}}) e^{(\mu V_0)^2 \frac{it}{2\mu}}  - \frac{1}{2} . \label{dCtinflimit}
\end{equation}

Instead of using Jordan-Wigner transformation, the discrete two-particle statevector $| k \rangle$'s can be mapped directly onto $\Gamma$ quantum registers.  The integrated correlation function in  Eq.(\ref{ctexpH}) can be rewritten as
\begin{equation}
C(t) =  \sum_{ k = \frac{2\pi n}{L}   }^{n \in [ - N, N]}    \langle 0 | \hat{P}_k  e^{- i \hat{H} t}  \hat{P}^\dag_k  |0 \rangle , \label{ctqcmap}
\end{equation}
where $N= \frac{2^\Gamma -1}{2}$ is cutoff of statevector $| k \rangle$'s, and  the projection operator $\hat{P}^\dag_k$ is defined by
\begin{equation}
 \hat{P}^\dag_k  = | k \rangle \langle 0 |.
\end{equation}
The Eq.(\ref{ctqcmap}) thus can be computed by using quantum circuit in Fig.~\ref{ctplot}. The unitary time evolution of kinematic term in Hamiltonian matrix in Eq.(\ref{Hmatkrep}) is simply given by a diagonal matrix,
\begin{equation} 
\langle k' | e^{ - i \hat{H}_0  t}  | k\rangle = \delta_{k, k'}  e^{ -  2 i \epsilon^{(0)}_k t}  .
\end{equation}
The second  potential term of in Hamiltonian matrix in Eq.(\ref{Hmatkrep}), $\hat{H}_V$,  is a constant matrix, hence we find
\begin{equation}
\langle k' | e^{ - i \hat{H}_V  t}  | k\rangle  = \begin{cases}  \frac{ e^{- i (2N+1) \frac{V_0}{L} t } + 2 N}{2N+1}, & \mbox{if} \ \ k=k' ; \\    \frac{ e^{- i (2N+1) \frac{V_0}{L} t } -1}{2N+1}, & \mbox{otherwise} . \end{cases}
\end{equation}
With $\Gamma$ quantum registers, the quantum circuit of $e^{ - i \hat{H}_V  t} $ is given by
\begin{align}
e^{ - i \hat{H}_V  t}  &= \frac{ e^{- i (2N+1) \frac{V_0}{L} t } + 2 N}{2N+1} I_{\Gamma -1} \otimes \cdots \otimes I_1 \otimes I_0 \nonumber \\
 &+ \frac{ e^{- i (2N+1) \frac{V_0}{L} t } -1}{2N+1} I_{\Gamma -1} \otimes \cdots \otimes I_1 \otimes X_0 \nonumber \\
 & + \cdots + \frac{ e^{- i (2N+1) \frac{V_0}{L} t } -1}{2N+1} X_{\Gamma -1} \otimes \cdots \otimes I_1 \otimes I_0 \nonumber \\
 &  + \frac{ e^{- i (2N+1) \frac{V_0}{L} t } -1}{2N+1} I_{\Gamma -1} \otimes \cdots \otimes X_1 \otimes X_0 \nonumber \\
 & + \cdots  + \frac{ e^{- i (2N+1) \frac{V_0}{L} t } -1}{2N+1} X_{\Gamma -1} \otimes \cdots \otimes X_1 \otimes X_0  ,
\end{align}
the sum include all the possible single $X$-gate insertions, two $X$-gate insertions, etc. up to $\Gamma$ $X$-gate insertion. The unitary time evolution of total Hamiltonian matrix then can be computed via trotterization approximation \cite{TrotterH.F.1959Otpo,Hatano:2005gh}, e.g. the lowest order of trotterziation is given by
\begin{equation}
e^{- i \hat{H} \delta t  } \stackrel{ \delta t \rightarrow 0}{\approx }   e^{- i  \hat{H}_{0} \delta t}   e^{- i  \hat{H}_{V} \delta t}     .
\end{equation}

\begin{figure}[h]
\includegraphics[width=0.85\textwidth]{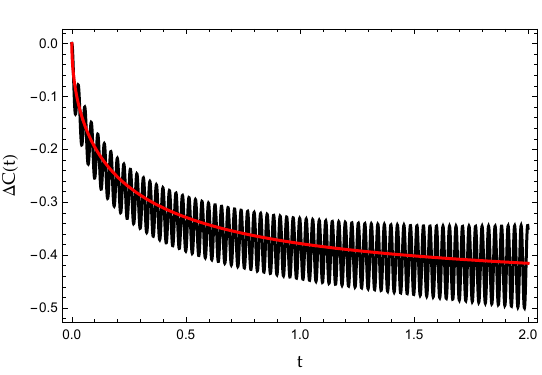}   
\caption{Demo plots of real part of  $ \triangle C(t) $ (black)  vs. its infinite volume limit given by the right-hand side of Eq.(\ref{dCtinflimit})  (red)  with parameters: $V_0=2.5$, $\mu =1$,  $L =  90$ and $N=300$. }
\label{redCtplot}
\end{figure}

\subsubsection{Data processing in real-time simulations}

\begin{figure}[h]
\includegraphics[width=0.85\textwidth]{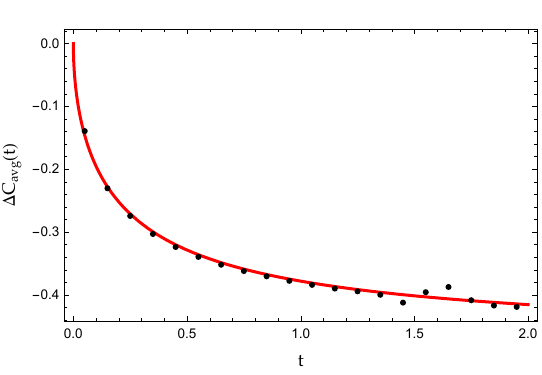}   
\caption{Demo plots of real part of  $ \triangle C_{avg}(t) $ (black)  vs. its infinite volume limit given by the right-hand side of Eq.(\ref{dCtinflimit})  (red)  with parameters: $V_0=2.5$, $\mu =1$,  $L =  90$, $N=1000$, $t_0=2$ and $N_t=20$. }
\label{redCtavgplot}
\end{figure}

The difference of integrated correlation functions, $\triangle C(t) = C (t) - C_0 (t)$, in the real time simulation is a fast oscillating function around its infinite volume limit, see demo plot in Fig.~\ref{redCtplot}.  The amplitude of fast oscillating function of $\triangle C(t)$ decrease slightly as momenta cutoff $N$ is increased. Meanwhile the frequency of oscillating motion increase as $N$ is increased, where the frequency of oscillation motion is proportional to momentum cutoff:  $\frac{2\pi N}{L}$.     We remark that the oscillatory behavior of correlation function for finite volume system with periodic boundary condition is in fact a quite well-known effect in nuclear physics, especially in finite volume scattering problems. For instance, the finite volume Green's function is typically periodic oscillating function compared with a smooth function of its infinite volume counterpart, see e.g.  discussion in \cite{Guo:2020ikh}.    The infinite limit of some finite volume quantities, such as Green's function and correlation function, can only be achieved with some careful mathematical manipulations,  such as  analytical continuation of finite volume Green's function into complex energy plane, see e.g.    \cite{Guo:2020ikh}.  The oscillating behavior can also be smoothed out via further data processing by averaging out $\triangle C(t)$ over a short period of time which is larger than oscillation period, see e.g. averaged tunneling time in a finite system as size of system is increased in Ref.~\cite{Guo:2022jyk}. Assuming the time of $\triangle C(t)$ is measured up to $t_0$,   let's split range of measurement up to $N_t$ segment of size of $\triangle t = \frac{t_0}{N_t}$ where $\triangle t  \gg \frac{L}{2\pi N}$,  the averaged $\triangle C(t)$  in $i$-th segment can be computed  by
\begin{equation}
\triangle C_{avg}(t_i) =  \frac{1}{\triangle t } \int_{t_{i-1}}^{t_i} \left [ C(t) -C_0 (t) \right ] d t, 
\end{equation}
where $i\in[1,N_t]$ and $t_i = (i - \frac{1}{2}) \triangle t $ sits at the center of $i$-th segment. The demo of   $\triangle C_{avg}(t_i)$ vs. its infinite volume limit is shown in Fig.~\ref{redCtavgplot}.

To extract the scattering phase shifts,  the scattering phase shift firstly must be modeled and parametrized   in terms of a few parameters and kinematic factors based on either chiral perturbation theory or K-matrix formalism, etc.  The phase shift can then be obtained by fitting quantum simulation data of integrated correlation functions with these free parameters through an integral.

\paragraph*{Summary and outlook.\textemdash} In summary, we propose to extract and determine the infinite volume scattering phase shifts through quantum simulation of the integrated correlation functions of few-particle systems in a trap on quantum computers. The formalism is based on the relation in Eq.(\ref{mainresult}) that relates  the integrated correlation functions for a trapped system to the infinite volume scattering phase shifts through a weighted integral in \cite{Guo:2023ecc,Guo:2024zal,Guo:2024pvt}. The Hamiltonian and correlation functions of two-particle system in the trap basis  can be easily mapped    onto quantum registers and are    suitable and convenient for quantum simulation. The integrated correlation function then can be computed by an ancilla-based algorithm proposed   in Ref.~\cite{PhysRevA.64.022319}, also see Fig.{\ref{ctplot}}. The proposal is demonstrated with a simple contact interaction fermion model that yields the analytic solutions in finite volume, and the further data processing to smooth out the fast oscillating behavior in real-time simulation is discussed.

\acknowledgments
This research was supported by the National Science Foundation under Grant No. NSF PHY-2418937 and in part by the National Science Foundation under Grant No. NSF PHY-2309135 to the Kavli Institute for Theoretical Physics (KITP).

\bibliography{ALL-REF.bib}

\begin{thebibliography}{61}%
\makeatletter
\providecommand \@ifxundefined [1]{%
 \@ifx{#1\undefined}
}%
\providecommand \@ifnum [1]{%
 \ifnum #1\expandafter \@firstoftwo
 \else \expandafter \@secondoftwo
 \fi
}%
\providecommand \@ifx [1]{%
 \ifx #1\expandafter \@firstoftwo
 \else \expandafter \@secondoftwo
 \fi
}%
\providecommand \natexlab [1]{#1}%
\providecommand \enquote  [1]{``#1''}%
\providecommand \bibnamefont  [1]{#1}%
\providecommand \bibfnamefont [1]{#1}%
\providecommand \citenamefont [1]{#1}%
\providecommand \href@noop [0]{\@secondoftwo}%
\providecommand \href [0]{\begingroup \@sanitize@url \@href}%
\providecommand \@href[1]{\@@startlink{#1}\@@href}%
\providecommand \@@href[1]{\endgroup#1\@@endlink}%
\providecommand \@sanitize@url [0]{\catcode `\\12\catcode `\$12\catcode
  `\&12\catcode `\#12\catcode `\^12\catcode `\_12\catcode `\%12\relax}%
\providecommand \@@startlink[1]{}%
\providecommand \@@endlink[0]{}%
\providecommand \url  [0]{\begingroup\@sanitize@url \@url }%
\providecommand \@url [1]{\endgroup\@href {#1}{\urlprefix }}%
\providecommand \urlprefix  [0]{URL }%
\providecommand \Eprint [0]{\href }%
\providecommand \doibase [0]{http://dx.doi.org/}%
\providecommand \selectlanguage [0]{\@gobble}%
\providecommand \bibinfo  [0]{\@secondoftwo}%
\providecommand \bibfield  [0]{\@secondoftwo}%
\providecommand \translation [1]{[#1]}%
\providecommand \BibitemOpen [0]{}%
\providecommand \bibitemStop [0]{}%
\providecommand \bibitemNoStop [0]{.\EOS\space}%
\providecommand \EOS [0]{\spacefactor3000\relax}%
\providecommand \BibitemShut  [1]{\csname bibitem#1\endcsname}%
\let\auto@bib@innerbib\@empty
\bibitem [{\citenamefont {Schaefer}\ \emph {et~al.}(2011)\citenamefont
  {Schaefer}, \citenamefont {Sommer},\ and\ \citenamefont
  {Virotta}}]{SCHAEFER201193}%
  \BibitemOpen
  \bibfield  {author} {\bibinfo {author} {\bibfnamefont {Stefan}\ \bibnamefont
  {Schaefer}}, \bibinfo {author} {\bibfnamefont {Rainer}\ \bibnamefont
  {Sommer}}, \ and\ \bibinfo {author} {\bibfnamefont {Francesco}\ \bibnamefont
  {Virotta}} (\bibinfo {collaboration} {ALPHA}),\ }\bibfield  {title} {\enquote
  {\bibinfo {title} {{Critical slowing down and error analysis in lattice QCD
  simulations}},}\ }\href {\doibase 10.1016/j.nuclphysb.2010.11.020} {\bibfield
   {journal} {\bibinfo  {journal} {Nucl. Phys. B}\ }\textbf {\bibinfo {volume}
  {845}},\ \bibinfo {pages} {93--119} (\bibinfo {year} {2011})},\ \Eprint
  {http://arxiv.org/abs/1009.5228} {arXiv:1009.5228 [hep-lat]} \BibitemShut
  {NoStop}%
\bibitem [{\citenamefont {Loh}\ \emph {et~al.}(1990)\citenamefont {Loh},
  \citenamefont {Gubernatis}, \citenamefont {Scalettar}, \citenamefont {White},
  \citenamefont {Scalapino},\ and\ \citenamefont {Sugar}}]{PhysRevB.41.9301}%
  \BibitemOpen
  \bibfield  {author} {\bibinfo {author} {\bibfnamefont {E.~Y.}\ \bibnamefont
  {Loh}}, \bibinfo {author} {\bibfnamefont {J.~E.}\ \bibnamefont {Gubernatis}},
  \bibinfo {author} {\bibfnamefont {R.~T.}\ \bibnamefont {Scalettar}}, \bibinfo
  {author} {\bibfnamefont {S.~R.}\ \bibnamefont {White}}, \bibinfo {author}
  {\bibfnamefont {D.~J.}\ \bibnamefont {Scalapino}}, \ and\ \bibinfo {author}
  {\bibfnamefont {R.~L.}\ \bibnamefont {Sugar}},\ }\bibfield  {title} {\enquote
  {\bibinfo {title} {{Sign problem in the numerical simulation of many-electron
  systems}},}\ }\href {\doibase 10.1103/PhysRevB.41.9301} {\bibfield  {journal}
  {\bibinfo  {journal} {Phys. Rev. B}\ }\textbf {\bibinfo {volume} {41}},\
  \bibinfo {pages} {9301--9307} (\bibinfo {year} {1990})}\BibitemShut {NoStop}%
\bibitem [{\citenamefont {de~Forcrand}(2009)}]{deForcrand:2009zkb}%
  \BibitemOpen
  \bibfield  {author} {\bibinfo {author} {\bibfnamefont {Philippe}\
  \bibnamefont {de~Forcrand}},\ }\bibfield  {title} {\enquote {\bibinfo {title}
  {{Simulating QCD at finite density}},}\ }\href {\doibase 10.22323/1.091.0010}
  {\bibfield  {journal} {\bibinfo  {journal} {PoS}\ }\textbf {\bibinfo {volume}
  {LAT2009}},\ \bibinfo {pages} {010} (\bibinfo {year} {2009})},\ \Eprint
  {http://arxiv.org/abs/1005.0539} {arXiv:1005.0539 [hep-lat]} \BibitemShut
  {NoStop}%
\bibitem [{\citenamefont {Lepage}(1989)}]{lepage1989analysis}%
  \BibitemOpen
  \bibfield  {author} {\bibinfo {author} {\bibfnamefont {G~Peter}\ \bibnamefont
  {Lepage}},\ }\bibfield  {title} {\enquote {\bibinfo {title} {The analysis of
  algorithms for lattice field theory},}\ }\href@noop {} {\bibfield  {journal}
  {\bibinfo  {journal} {Boulder ASI}\ }\textbf {\bibinfo {volume} {1989}},\
  \bibinfo {pages} {97--120} (\bibinfo {year} {1989})}\BibitemShut {NoStop}%
\bibitem [{\citenamefont {Drischler}\ \emph {et~al.}(2021)\citenamefont
  {Drischler}, \citenamefont {Haxton}, \citenamefont {McElvain}, \citenamefont
  {Mereghetti}, \citenamefont {Nicholson}, \citenamefont {Vranas},\ and\
  \citenamefont {Walker-Loud}}]{DRISCHLER2021103888}%
  \BibitemOpen
  \bibfield  {author} {\bibinfo {author} {\bibfnamefont {Christian}\
  \bibnamefont {Drischler}}, \bibinfo {author} {\bibfnamefont {Wick}\
  \bibnamefont {Haxton}}, \bibinfo {author} {\bibfnamefont {Kenneth}\
  \bibnamefont {McElvain}}, \bibinfo {author} {\bibfnamefont {Emanuele}\
  \bibnamefont {Mereghetti}}, \bibinfo {author} {\bibfnamefont {Amy}\
  \bibnamefont {Nicholson}}, \bibinfo {author} {\bibfnamefont {Pavlos}\
  \bibnamefont {Vranas}}, \ and\ \bibinfo {author} {\bibfnamefont {André}\
  \bibnamefont {Walker-Loud}},\ }\bibfield  {title} {\enquote {\bibinfo {title}
  {Towards grounding nuclear physics in qcd},}\ }\href {\doibase
  https://doi.org/10.1016/j.ppnp.2021.103888} {\bibfield  {journal} {\bibinfo
  {journal} {Progress in Particle and Nuclear Physics}\ }\textbf {\bibinfo
  {volume} {121}},\ \bibinfo {pages} {103888} (\bibinfo {year}
  {2021})}\BibitemShut {NoStop}%
\bibitem [{\citenamefont {L{\"u}scher}(1991)}]{Luscher:1990ux}%
  \BibitemOpen
  \bibfield  {author} {\bibinfo {author} {\bibfnamefont {Martin}\ \bibnamefont
  {L{\"u}scher}},\ }\bibfield  {title} {\enquote {\bibinfo {title} {{Two
  particle states on a torus and their relation to the scattering matrix}},}\
  }\href {\doibase 10.1016/0550-3213(91)90366-6} {\bibfield  {journal}
  {\bibinfo  {journal} {Nucl. Phys.}\ }\textbf {\bibinfo {volume} {B354}},\
  \bibinfo {pages} {531--578} (\bibinfo {year} {1991})}\BibitemShut {NoStop}%
\bibitem [{\citenamefont {Busch}\ \emph {et~al.}(1998)\citenamefont {Busch},
  \citenamefont {Englert}, \citenamefont {Rza\.zewski},\ and\ \citenamefont
  {Wilkens}}]{Busch98}%
  \BibitemOpen
  \bibfield  {author} {\bibinfo {author} {\bibfnamefont {T.}~\bibnamefont
  {Busch}}, \bibinfo {author} {\bibfnamefont {B.-G.}\ \bibnamefont {Englert}},
  \bibinfo {author} {\bibfnamefont {K.}~\bibnamefont {Rza\.zewski}}, \ and\
  \bibinfo {author} {\bibfnamefont {M.}~\bibnamefont {Wilkens}},\ }\bibfield
  {title} {\enquote {\bibinfo {title} {{Two Cold Atoms in a Harmonic Trap}},}\
  }\href {\doibase 10.1023/A:1018705520999} {\bibfield  {journal} {\bibinfo
  {journal} {Found. Phys.}\ }\textbf {\bibinfo {volume} {28}},\ \bibinfo
  {pages} {549–559} (\bibinfo {year} {1998})}\BibitemShut {NoStop}%
\bibitem [{\citenamefont {Rummukainen}\ and\ \citenamefont
  {Gottlieb}(1995)}]{Rummukainen:1995vs}%
  \BibitemOpen
  \bibfield  {author} {\bibinfo {author} {\bibfnamefont {K.}~\bibnamefont
  {Rummukainen}}\ and\ \bibinfo {author} {\bibfnamefont {Steven~A.}\
  \bibnamefont {Gottlieb}},\ }\bibfield  {title} {\enquote {\bibinfo {title}
  {{Resonance scattering phase shifts on a nonrest frame lattice}},}\ }\href
  {\doibase 10.1016/0550-3213(95)00313-H} {\bibfield  {journal} {\bibinfo
  {journal} {Nucl. Phys.}\ }\textbf {\bibinfo {volume} {B450}},\ \bibinfo
  {pages} {397--436} (\bibinfo {year} {1995})},\ \Eprint
  {http://arxiv.org/abs/hep-lat/9503028} {arXiv:hep-lat/9503028 [hep-lat]}
  \BibitemShut {NoStop}%
\bibitem [{\citenamefont {Christ}\ \emph {et~al.}(2005)\citenamefont {Christ},
  \citenamefont {Kim},\ and\ \citenamefont {Yamazaki}}]{Christ:2005gi}%
  \BibitemOpen
  \bibfield  {author} {\bibinfo {author} {\bibfnamefont {Norman~H.}\
  \bibnamefont {Christ}}, \bibinfo {author} {\bibfnamefont {Changhoan}\
  \bibnamefont {Kim}}, \ and\ \bibinfo {author} {\bibfnamefont {Takeshi}\
  \bibnamefont {Yamazaki}},\ }\bibfield  {title} {\enquote {\bibinfo {title}
  {{Finite volume corrections to the two-particle decay of states with non-zero
  momentum}},}\ }\href {\doibase 10.1103/PhysRevD.72.114506} {\bibfield
  {journal} {\bibinfo  {journal} {Phys. Rev.}\ }\textbf {\bibinfo {volume}
  {D72}},\ \bibinfo {pages} {114506} (\bibinfo {year} {2005})},\ \Eprint
  {http://arxiv.org/abs/hep-lat/0507009} {arXiv:hep-lat/0507009 [hep-lat]}
  \BibitemShut {NoStop}%
\bibitem [{\citenamefont {Bernard}\ \emph {et~al.}(2008)\citenamefont
  {Bernard}, \citenamefont {Lage}, \citenamefont {Mei{\ss}ner},\ and\
  \citenamefont {Rusetsky}}]{Bernard:2008ax}%
  \BibitemOpen
  \bibfield  {author} {\bibinfo {author} {\bibfnamefont {V.}~\bibnamefont
  {Bernard}}, \bibinfo {author} {\bibfnamefont {M.}~\bibnamefont {Lage}},
  \bibinfo {author} {\bibfnamefont {U.-G.}\ \bibnamefont {Mei{\ss}ner}}, \ and\
  \bibinfo {author} {\bibfnamefont {A.}~\bibnamefont {Rusetsky}},\ }\bibfield
  {title} {\enquote {\bibinfo {title} {{Resonance properties from the
  finite-volume energy spectrum}},}\ }\href {\doibase
  10.1088/1126-6708/2008/08/024} {\bibfield  {journal} {\bibinfo  {journal}
  {JHEP}\ }\textbf {\bibinfo {volume} {08}},\ \bibinfo {pages} {024} (\bibinfo
  {year} {2008})},\ \Eprint {http://arxiv.org/abs/0806.4495} {arXiv:0806.4495
  [hep-lat]} \BibitemShut {NoStop}%
\bibitem [{\citenamefont {He}\ \emph {et~al.}(2005)\citenamefont {He},
  \citenamefont {Feng},\ and\ \citenamefont {Liu}}]{He:2005ey}%
  \BibitemOpen
  \bibfield  {author} {\bibinfo {author} {\bibfnamefont {Song}\ \bibnamefont
  {He}}, \bibinfo {author} {\bibfnamefont {Xu}~\bibnamefont {Feng}}, \ and\
  \bibinfo {author} {\bibfnamefont {Chuan}\ \bibnamefont {Liu}},\ }\bibfield
  {title} {\enquote {\bibinfo {title} {{Two particle states and the S-matrix
  elements in multi-channel scattering}},}\ }\href {\doibase
  10.1088/1126-6708/2005/07/011} {\bibfield  {journal} {\bibinfo  {journal}
  {JHEP}\ }\textbf {\bibinfo {volume} {07}},\ \bibinfo {pages} {011} (\bibinfo
  {year} {2005})},\ \Eprint {http://arxiv.org/abs/hep-lat/0504019}
  {arXiv:hep-lat/0504019 [hep-lat]} \BibitemShut {NoStop}%
\bibitem [{\citenamefont {Lage}\ \emph {et~al.}(2009)\citenamefont {Lage},
  \citenamefont {Mei{\ss}ner},\ and\ \citenamefont {Rusetsky}}]{Lage:2009zv}%
  \BibitemOpen
  \bibfield  {author} {\bibinfo {author} {\bibfnamefont {Michael}\ \bibnamefont
  {Lage}}, \bibinfo {author} {\bibfnamefont {Ulf-G.}\ \bibnamefont
  {Mei{\ss}ner}}, \ and\ \bibinfo {author} {\bibfnamefont {Akaki}\ \bibnamefont
  {Rusetsky}},\ }\bibfield  {title} {\enquote {\bibinfo {title} {{A Method to
  measure the antikaon-nucleon scattering length in lattice QCD}},}\ }\href
  {\doibase 10.1016/j.physletb.2009.10.055} {\bibfield  {journal} {\bibinfo
  {journal} {Phys. Lett.}\ }\textbf {\bibinfo {volume} {B681}},\ \bibinfo
  {pages} {439--443} (\bibinfo {year} {2009})},\ \Eprint
  {http://arxiv.org/abs/0905.0069} {arXiv:0905.0069 [hep-lat]} \BibitemShut
  {NoStop}%
\bibitem [{\citenamefont {D{\"o}ring}\ \emph {et~al.}(2011)\citenamefont
  {D{\"o}ring}, \citenamefont {Mei{\ss}ner}, \citenamefont {Oset},\ and\
  \citenamefont {Rusetsky}}]{Doring:2011vk}%
  \BibitemOpen
  \bibfield  {author} {\bibinfo {author} {\bibfnamefont {M.}~\bibnamefont
  {D{\"o}ring}}, \bibinfo {author} {\bibfnamefont {U.-G.}\ \bibnamefont
  {Mei{\ss}ner}}, \bibinfo {author} {\bibfnamefont {E.}~\bibnamefont {Oset}}, \
  and\ \bibinfo {author} {\bibfnamefont {A.}~\bibnamefont {Rusetsky}},\
  }\bibfield  {title} {\enquote {\bibinfo {title} {{Unitarized Chiral
  Perturbation Theory in a finite volume: Scalar meson sector}},}\ }\href
  {\doibase 10.1140/epja/i2011-11139-7} {\bibfield  {journal} {\bibinfo
  {journal} {Eur. Phys. J.}\ }\textbf {\bibinfo {volume} {A47}},\ \bibinfo
  {pages} {139} (\bibinfo {year} {2011})},\ \Eprint
  {http://arxiv.org/abs/1107.3988} {arXiv:1107.3988 [hep-lat]} \BibitemShut
  {NoStop}%
\bibitem [{\citenamefont {Guo}\ \emph {et~al.}(2013)\citenamefont {Guo},
  \citenamefont {Dudek}, \citenamefont {Edwards},\ and\ \citenamefont
  {Szczepaniak}}]{Guo:2012hv}%
  \BibitemOpen
  \bibfield  {author} {\bibinfo {author} {\bibfnamefont {Peng}\ \bibnamefont
  {Guo}}, \bibinfo {author} {\bibfnamefont {Jozef}\ \bibnamefont {Dudek}},
  \bibinfo {author} {\bibfnamefont {Robert}\ \bibnamefont {Edwards}}, \ and\
  \bibinfo {author} {\bibfnamefont {Adam~P.}\ \bibnamefont {Szczepaniak}},\
  }\bibfield  {title} {\enquote {\bibinfo {title} {{Coupled-channel scattering
  on a torus}},}\ }\href {\doibase 10.1103/PhysRevD.88.014501} {\bibfield
  {journal} {\bibinfo  {journal} {Phys. Rev.}\ }\textbf {\bibinfo {volume}
  {D88}},\ \bibinfo {pages} {014501} (\bibinfo {year} {2013})},\ \Eprint
  {http://arxiv.org/abs/1211.0929} {arXiv:1211.0929 [hep-lat]} \BibitemShut
  {NoStop}%
\bibitem [{\citenamefont {Guo}(2013)}]{Guo:2013vsa}%
  \BibitemOpen
  \bibfield  {author} {\bibinfo {author} {\bibfnamefont {Peng}\ \bibnamefont
  {Guo}},\ }\bibfield  {title} {\enquote {\bibinfo {title} {{Coupled-channel
  scattering in 1+1 dimensional lattice model}},}\ }\href {\doibase
  10.1103/PhysRevD.88.014507} {\bibfield  {journal} {\bibinfo  {journal} {Phys.
  Rev.}\ }\textbf {\bibinfo {volume} {D88}},\ \bibinfo {pages} {014507}
  (\bibinfo {year} {2013})},\ \Eprint {http://arxiv.org/abs/1304.7812}
  {arXiv:1304.7812 [hep-lat]} \BibitemShut {NoStop}%
\bibitem [{\citenamefont {Kreuzer}\ and\ \citenamefont
  {Hammer}(2009)}]{Kreuzer:2008bi}%
  \BibitemOpen
  \bibfield  {author} {\bibinfo {author} {\bibfnamefont {Simon}\ \bibnamefont
  {Kreuzer}}\ and\ \bibinfo {author} {\bibfnamefont {H.~W.}\ \bibnamefont
  {Hammer}},\ }\bibfield  {title} {\enquote {\bibinfo {title} {{Efimov physics
  in a finite volume}},}\ }\href {\doibase 10.1016/j.physletb.2009.02.035}
  {\bibfield  {journal} {\bibinfo  {journal} {Phys. Lett.}\ }\textbf {\bibinfo
  {volume} {B673}},\ \bibinfo {pages} {260--263} (\bibinfo {year} {2009})},\
  \Eprint {http://arxiv.org/abs/0811.0159} {arXiv:0811.0159 [nucl-th]}
  \BibitemShut {NoStop}%
\bibitem [{\citenamefont {Polejaeva}\ and\ \citenamefont
  {Rusetsky}(2012)}]{Polejaeva:2012ut}%
  \BibitemOpen
  \bibfield  {author} {\bibinfo {author} {\bibfnamefont {K.}~\bibnamefont
  {Polejaeva}}\ and\ \bibinfo {author} {\bibfnamefont {A.}~\bibnamefont
  {Rusetsky}},\ }\bibfield  {title} {\enquote {\bibinfo {title} {{Three
  particles in a finite volume}},}\ }\href {\doibase
  10.1140/epja/i2012-12067-8} {\bibfield  {journal} {\bibinfo  {journal} {Eur.
  Phys. J.}\ }\textbf {\bibinfo {volume} {A48}},\ \bibinfo {pages} {67}
  (\bibinfo {year} {2012})},\ \Eprint {http://arxiv.org/abs/1203.1241}
  {arXiv:1203.1241 [hep-lat]} \BibitemShut {NoStop}%
\bibitem [{\citenamefont {Hansen}\ and\ \citenamefont
  {Sharpe}(2014)}]{Hansen:2014eka}%
  \BibitemOpen
  \bibfield  {author} {\bibinfo {author} {\bibfnamefont {Maxwell~T.}\
  \bibnamefont {Hansen}}\ and\ \bibinfo {author} {\bibfnamefont {Stephen~R.}\
  \bibnamefont {Sharpe}},\ }\bibfield  {title} {\enquote {\bibinfo {title}
  {{Relativistic, model-independent, three-particle quantization condition}},}\
  }\href {\doibase 10.1103/PhysRevD.90.116003} {\bibfield  {journal} {\bibinfo
  {journal} {Phys. Rev.}\ }\textbf {\bibinfo {volume} {D90}},\ \bibinfo {pages}
  {116003} (\bibinfo {year} {2014})},\ \Eprint {http://arxiv.org/abs/1408.5933}
  {arXiv:1408.5933 [hep-lat]} \BibitemShut {NoStop}%
\bibitem [{\citenamefont {Mai}\ and\ \citenamefont
  {D{\"o}ring}(2017)}]{Mai:2017bge}%
  \BibitemOpen
  \bibfield  {author} {\bibinfo {author} {\bibfnamefont {M.}~\bibnamefont
  {Mai}}\ and\ \bibinfo {author} {\bibfnamefont {M.}~\bibnamefont
  {D{\"o}ring}},\ }\bibfield  {title} {\enquote {\bibinfo {title} {{Three-body
  Unitarity in the Finite Volume}},}\ }\href {\doibase
  10.1140/epja/i2017-12440-1} {\bibfield  {journal} {\bibinfo  {journal} {Eur.
  Phys. J.}\ }\textbf {\bibinfo {volume} {A53}},\ \bibinfo {pages} {240}
  (\bibinfo {year} {2017})},\ \Eprint {http://arxiv.org/abs/1709.08222}
  {arXiv:1709.08222 [hep-lat]} \BibitemShut {NoStop}%
\bibitem [{\citenamefont {Mai}\ and\ \citenamefont
  {Döring}(2019)}]{Mai:2018djl}%
  \BibitemOpen
  \bibfield  {author} {\bibinfo {author} {\bibfnamefont {Maxim}\ \bibnamefont
  {Mai}}\ and\ \bibinfo {author} {\bibfnamefont {Michael}\ \bibnamefont
  {Döring}},\ }\bibfield  {title} {\enquote {\bibinfo {title} {{Finite-Volume
  Spectrum of $\pi^+\pi^+$ and $\pi^+\pi^+\pi^+$ Systems}},}\ }\href {\doibase
  10.1103/PhysRevLett.122.062503} {\bibfield  {journal} {\bibinfo  {journal}
  {Phys. Rev. Lett.}\ }\textbf {\bibinfo {volume} {122}},\ \bibinfo {pages}
  {062503} (\bibinfo {year} {2019})},\ \Eprint
  {http://arxiv.org/abs/1807.04746} {arXiv:1807.04746 [hep-lat]} \BibitemShut
  {NoStop}%
\bibitem [{\citenamefont {D{\"o}ring}\ \emph {et~al.}(2018)\citenamefont
  {D{\"o}ring}, \citenamefont {Hammer}, \citenamefont {Mai}, \citenamefont
  {Pang}, \citenamefont {Rusetsky},\ and\ \citenamefont {Wu}}]{Doring:2018xxx}%
  \BibitemOpen
  \bibfield  {author} {\bibinfo {author} {\bibfnamefont {M.}~\bibnamefont
  {D{\"o}ring}}, \bibinfo {author} {\bibfnamefont {H.~W.}\ \bibnamefont
  {Hammer}}, \bibinfo {author} {\bibfnamefont {M.}~\bibnamefont {Mai}},
  \bibinfo {author} {\bibfnamefont {J.~Y}\ \bibnamefont {Pang}}, \bibinfo
  {author} {\bibfnamefont {A.}~\bibnamefont {Rusetsky}}, \ and\ \bibinfo
  {author} {\bibfnamefont {J.}~\bibnamefont {Wu}},\ }\bibfield  {title}
  {\enquote {\bibinfo {title} {{Three-body spectrum in a finite volume: the
  role of cubic symmetry}},}\ }\href {\doibase 10.1103/PhysRevD.97.114508}
  {\bibfield  {journal} {\bibinfo  {journal} {Phys. Rev.}\ }\textbf {\bibinfo
  {volume} {D97}},\ \bibinfo {pages} {114508} (\bibinfo {year} {2018})},\
  \Eprint {http://arxiv.org/abs/1802.03362} {arXiv:1802.03362 [hep-lat]}
  \BibitemShut {NoStop}%
\bibitem [{\citenamefont {Guo}(2017)}]{Guo:2016fgl}%
  \BibitemOpen
  \bibfield  {author} {\bibinfo {author} {\bibfnamefont {Peng}\ \bibnamefont
  {Guo}},\ }\bibfield  {title} {\enquote {\bibinfo {title} {{One spatial
  dimensional finite volume three-body interaction for a short-range
  potential}},}\ }\href {\doibase 10.1103/PhysRevD.95.054508} {\bibfield
  {journal} {\bibinfo  {journal} {Phys. Rev.}\ }\textbf {\bibinfo {volume}
  {D95}},\ \bibinfo {pages} {054508} (\bibinfo {year} {2017})},\ \Eprint
  {http://arxiv.org/abs/1607.03184} {arXiv:1607.03184 [hep-lat]} \BibitemShut
  {NoStop}%
\bibitem [{\citenamefont {Guo}\ and\ \citenamefont
  {Gasparian}(2017)}]{Guo:2017ism}%
  \BibitemOpen
  \bibfield  {author} {\bibinfo {author} {\bibfnamefont {Peng}\ \bibnamefont
  {Guo}}\ and\ \bibinfo {author} {\bibfnamefont {Vladimir}\ \bibnamefont
  {Gasparian}},\ }\bibfield  {title} {\enquote {\bibinfo {title} {{An solvable
  three-body model in finite volume}},}\ }\href {\doibase
  10.1016/j.physletb.2017.10.009} {\bibfield  {journal} {\bibinfo  {journal}
  {Phys. Lett.}\ }\textbf {\bibinfo {volume} {B774}},\ \bibinfo {pages}
  {441--445} (\bibinfo {year} {2017})},\ \Eprint
  {http://arxiv.org/abs/1701.00438} {arXiv:1701.00438 [hep-lat]} \BibitemShut
  {NoStop}%
\bibitem [{\citenamefont {Guo}\ and\ \citenamefont
  {Gasparian}(2018)}]{Guo:2017crd}%
  \BibitemOpen
  \bibfield  {author} {\bibinfo {author} {\bibfnamefont {Peng}\ \bibnamefont
  {Guo}}\ and\ \bibinfo {author} {\bibfnamefont {Vladimir}\ \bibnamefont
  {Gasparian}},\ }\bibfield  {title} {\enquote {\bibinfo {title} {{Numerical
  approach for finite volume three-body interaction}},}\ }\href {\doibase
  10.1103/PhysRevD.97.014504} {\bibfield  {journal} {\bibinfo  {journal} {Phys.
  Rev.}\ }\textbf {\bibinfo {volume} {D97}},\ \bibinfo {pages} {014504}
  (\bibinfo {year} {2018})},\ \Eprint {http://arxiv.org/abs/1709.08255}
  {arXiv:1709.08255 [hep-lat]} \BibitemShut {NoStop}%
\bibitem [{\citenamefont {Guo}\ and\ \citenamefont
  {Morris}(2019)}]{Guo:2018xbv}%
  \BibitemOpen
  \bibfield  {author} {\bibinfo {author} {\bibfnamefont {Peng}\ \bibnamefont
  {Guo}}\ and\ \bibinfo {author} {\bibfnamefont {Tyler}\ \bibnamefont
  {Morris}},\ }\bibfield  {title} {\enquote {\bibinfo {title}
  {{Multiple-particle interaction in (1+1)-dimensional lattice model}},}\
  }\href {\doibase 10.1103/PhysRevD.99.014501} {\bibfield  {journal} {\bibinfo
  {journal} {Phys. Rev.}\ }\textbf {\bibinfo {volume} {D99}},\ \bibinfo {pages}
  {014501} (\bibinfo {year} {2019})},\ \Eprint
  {http://arxiv.org/abs/1808.07397} {arXiv:1808.07397 [hep-lat]} \BibitemShut
  {NoStop}%
\bibitem [{\citenamefont {Mai}\ \emph {et~al.}(2020)\citenamefont {Mai},
  \citenamefont {D\"oring}, \citenamefont {Culver},\ and\ \citenamefont
  {Alexandru}}]{Mai:2019fba}%
  \BibitemOpen
  \bibfield  {author} {\bibinfo {author} {\bibfnamefont {M.}~\bibnamefont
  {Mai}}, \bibinfo {author} {\bibfnamefont {M.}~\bibnamefont {D\"oring}},
  \bibinfo {author} {\bibfnamefont {C.}~\bibnamefont {Culver}}, \ and\ \bibinfo
  {author} {\bibfnamefont {A.}~\bibnamefont {Alexandru}},\ }\bibfield  {title}
  {\enquote {\bibinfo {title} {{Three-body unitarity versus finite-volume
  $\pi^+\pi^+\pi^+$ spectrum from lattice QCD}},}\ }\href {\doibase
  10.1103/PhysRevD.101.054510} {\bibfield  {journal} {\bibinfo  {journal}
  {Phys. Rev. D}\ }\textbf {\bibinfo {volume} {101}},\ \bibinfo {pages}
  {054510} (\bibinfo {year} {2020})},\ \Eprint
  {http://arxiv.org/abs/1909.05749} {arXiv:1909.05749 [hep-lat]} \BibitemShut
  {NoStop}%
\bibitem [{\citenamefont {Guo}\ \emph {et~al.}(2018)\citenamefont {Guo},
  \citenamefont {Döring},\ and\ \citenamefont {Szczepaniak}}]{Guo:2018ibd}%
  \BibitemOpen
  \bibfield  {author} {\bibinfo {author} {\bibfnamefont {Peng}\ \bibnamefont
  {Guo}}, \bibinfo {author} {\bibfnamefont {Michael}\ \bibnamefont {Döring}},
  \ and\ \bibinfo {author} {\bibfnamefont {Adam~P.}\ \bibnamefont
  {Szczepaniak}},\ }\bibfield  {title} {\enquote {\bibinfo {title}
  {{Variational approach to $N$-body interactions in finite volume}},}\ }\href
  {\doibase 10.1103/PhysRevD.98.094502} {\bibfield  {journal} {\bibinfo
  {journal} {Phys. Rev.}\ }\textbf {\bibinfo {volume} {D98}},\ \bibinfo {pages}
  {094502} (\bibinfo {year} {2018})},\ \Eprint
  {http://arxiv.org/abs/1810.01261} {arXiv:1810.01261 [hep-lat]} \BibitemShut
  {NoStop}%
\bibitem [{\citenamefont {Guo}(2020{\natexlab{a}})}]{Guo:2019hih}%
  \BibitemOpen
  \bibfield  {author} {\bibinfo {author} {\bibfnamefont {Peng}\ \bibnamefont
  {Guo}},\ }\bibfield  {title} {\enquote {\bibinfo {title} {{Propagation of
  particles on a torus}},}\ }\href {\doibase 10.1016/j.physletb.2020.135370}
  {\bibfield  {journal} {\bibinfo  {journal} {Phys. Lett. B}\ }\textbf
  {\bibinfo {volume} {804}},\ \bibinfo {pages} {135370} (\bibinfo {year}
  {2020}{\natexlab{a}})},\ \Eprint {http://arxiv.org/abs/1908.08081}
  {arXiv:1908.08081 [hep-lat]} \BibitemShut {NoStop}%
\bibitem [{\citenamefont {Guo}\ and\ \citenamefont
  {D\"oring}(2020)}]{Guo:2019ogp}%
  \BibitemOpen
  \bibfield  {author} {\bibinfo {author} {\bibfnamefont {Peng}\ \bibnamefont
  {Guo}}\ and\ \bibinfo {author} {\bibfnamefont {Michael}\ \bibnamefont
  {D\"oring}},\ }\bibfield  {title} {\enquote {\bibinfo {title} {{Lattice model
  of heavy-light three-body system}},}\ }\href {\doibase
  10.1103/PhysRevD.101.034501} {\bibfield  {journal} {\bibinfo  {journal}
  {Phys. Rev. D}\ }\textbf {\bibinfo {volume} {101}},\ \bibinfo {pages}
  {034501} (\bibinfo {year} {2020})},\ \Eprint
  {http://arxiv.org/abs/1910.08624} {arXiv:1910.08624 [hep-lat]} \BibitemShut
  {NoStop}%
\bibitem [{\citenamefont {Guo}(2020{\natexlab{b}})}]{Guo:2020wbl}%
  \BibitemOpen
  \bibfield  {author} {\bibinfo {author} {\bibfnamefont {Peng}\ \bibnamefont
  {Guo}},\ }\bibfield  {title} {\enquote {\bibinfo {title} {{Threshold
  expansion formula of $N$ bosons in a finite volume from a variational
  approach}},}\ }\href {\doibase 10.1103/PhysRevD.101.054512} {\bibfield
  {journal} {\bibinfo  {journal} {Phys. Rev. D}\ }\textbf {\bibinfo {volume}
  {101}},\ \bibinfo {pages} {054512} (\bibinfo {year} {2020}{\natexlab{b}})},\
  \Eprint {http://arxiv.org/abs/2002.04111} {arXiv:2002.04111 [hep-lat]}
  \BibitemShut {NoStop}%
\bibitem [{\citenamefont {Guo}\ and\ \citenamefont
  {Long}(2020{\natexlab{a}})}]{Guo:2020kph}%
  \BibitemOpen
  \bibfield  {author} {\bibinfo {author} {\bibfnamefont {Peng}\ \bibnamefont
  {Guo}}\ and\ \bibinfo {author} {\bibfnamefont {Bingwei}\ \bibnamefont
  {Long}},\ }\bibfield  {title} {\enquote {\bibinfo {title} {{Multi- $\pi^+$
  systems in a finite volume}},}\ }\href {\doibase 10.1103/PhysRevD.101.094510}
  {\bibfield  {journal} {\bibinfo  {journal} {Phys. Rev. D}\ }\textbf {\bibinfo
  {volume} {101}},\ \bibinfo {pages} {094510} (\bibinfo {year}
  {2020}{\natexlab{a}})},\ \Eprint {http://arxiv.org/abs/2002.09266}
  {arXiv:2002.09266 [hep-lat]} \BibitemShut {NoStop}%
\bibitem [{\citenamefont {Guo}(2020{\natexlab{c}})}]{Guo:2020iep}%
  \BibitemOpen
  \bibfield  {author} {\bibinfo {author} {\bibfnamefont {Peng}\ \bibnamefont
  {Guo}},\ }\bibfield  {title} {\enquote {\bibinfo {title} {{Myth of scattering
  in finite volume}},}\ }\href@noop {} {\  (\bibinfo {year}
  {2020}{\natexlab{c}})},\ \Eprint {http://arxiv.org/abs/2007.04473}
  {arXiv:2007.04473 [hep-lat]} \BibitemShut {NoStop}%
\bibitem [{\citenamefont {Guo}\ and\ \citenamefont
  {Long}(2020{\natexlab{b}})}]{Guo:2020ikh}%
  \BibitemOpen
  \bibfield  {author} {\bibinfo {author} {\bibfnamefont {Peng}\ \bibnamefont
  {Guo}}\ and\ \bibinfo {author} {\bibfnamefont {Bingwei}\ \bibnamefont
  {Long}},\ }\bibfield  {title} {\enquote {\bibinfo {title} {{Visualizing
  resonances in finite volume}},}\ }\href {\doibase
  10.1103/PhysRevD.102.074508} {\bibfield  {journal} {\bibinfo  {journal}
  {Phys. Rev. D}\ }\textbf {\bibinfo {volume} {102}},\ \bibinfo {pages}
  {074508} (\bibinfo {year} {2020}{\natexlab{b}})},\ \Eprint
  {http://arxiv.org/abs/2007.10895} {arXiv:2007.10895 [hep-lat]} \BibitemShut
  {NoStop}%
\bibitem [{\citenamefont {Guo}(2020{\natexlab{d}})}]{Guo:2020spn}%
  \BibitemOpen
  \bibfield  {author} {\bibinfo {author} {\bibfnamefont {Peng}\ \bibnamefont
  {Guo}},\ }\bibfield  {title} {\enquote {\bibinfo {title} {{Modeling few-body
  resonances in finite volume}},}\ }\href {\doibase
  10.1103/PhysRevD.102.054514} {\bibfield  {journal} {\bibinfo  {journal}
  {Phys. Rev. D}\ }\textbf {\bibinfo {volume} {102}},\ \bibinfo {pages}
  {054514} (\bibinfo {year} {2020}{\natexlab{d}})},\ \Eprint
  {http://arxiv.org/abs/2007.12790} {arXiv:2007.12790 [hep-lat]} \BibitemShut
  {NoStop}%
\bibitem [{\citenamefont {Guo}\ and\ \citenamefont
  {Gasparian}(2021)}]{Guo:2021lhz}%
  \BibitemOpen
  \bibfield  {author} {\bibinfo {author} {\bibfnamefont {Peng}\ \bibnamefont
  {Guo}}\ and\ \bibinfo {author} {\bibfnamefont {Vladimir}\ \bibnamefont
  {Gasparian}},\ }\bibfield  {title} {\enquote {\bibinfo {title} {{Charged
  particles interaction in both a finite volume and a uniform magnetic
  field}},}\ }\href {\doibase 10.1103/PhysRevD.103.094520} {\bibfield
  {journal} {\bibinfo  {journal} {Phys. Rev. D}\ }\textbf {\bibinfo {volume}
  {103}},\ \bibinfo {pages} {094520} (\bibinfo {year} {2021})},\ \Eprint
  {http://arxiv.org/abs/2101.01150} {arXiv:2101.01150 [hep-lat]} \BibitemShut
  {NoStop}%
\bibitem [{\citenamefont {Guo}\ and\ \citenamefont {Long}(2022)}]{Guo:2021uig}%
  \BibitemOpen
  \bibfield  {author} {\bibinfo {author} {\bibfnamefont {Peng}\ \bibnamefont
  {Guo}}\ and\ \bibinfo {author} {\bibfnamefont {Bingwei}\ \bibnamefont
  {Long}},\ }\bibfield  {title} {\enquote {\bibinfo {title} {{Nuclear reactions
  in artificial traps}},}\ }\href {\doibase 10.1088/1361-6471/ac59d5}
  {\bibfield  {journal} {\bibinfo  {journal} {J. Phys. G}\ }\textbf {\bibinfo
  {volume} {49}},\ \bibinfo {pages} {055104} (\bibinfo {year} {2022})},\
  \Eprint {http://arxiv.org/abs/2101.03901} {arXiv:2101.03901 [nucl-th]}
  \BibitemShut {NoStop}%
\bibitem [{\citenamefont {Guo}(2021)}]{Guo:2021qfu}%
  \BibitemOpen
  \bibfield  {author} {\bibinfo {author} {\bibfnamefont {Peng}\ \bibnamefont
  {Guo}},\ }\bibfield  {title} {\enquote {\bibinfo {title} {{Coulomb
  corrections to two-particle interactions in artificial traps}},}\ }\href
  {\doibase 10.1103/PhysRevC.103.064611} {\bibfield  {journal} {\bibinfo
  {journal} {Phys. Rev. C}\ }\textbf {\bibinfo {volume} {103}},\ \bibinfo
  {pages} {064611} (\bibinfo {year} {2021})},\ \Eprint
  {http://arxiv.org/abs/2101.11097} {arXiv:2101.11097 [nucl-th]} \BibitemShut
  {NoStop}%
\bibitem [{\citenamefont {Guo}\ and\ \citenamefont
  {Gasparian}(2022)}]{Guo:2021hrf}%
  \BibitemOpen
  \bibfield  {author} {\bibinfo {author} {\bibfnamefont {Peng}\ \bibnamefont
  {Guo}}\ and\ \bibinfo {author} {\bibfnamefont {Vladimir}\ \bibnamefont
  {Gasparian}},\ }\bibfield  {title} {\enquote {\bibinfo {title} {{Charged
  particles interaction in both a finite volume and a uniform magnetic field
  II: topological and analytic properties of a magnetic system}},}\ }\href
  {\doibase 10.1088/1751-8121/ac7180} {\bibfield  {journal} {\bibinfo
  {journal} {J. Phys. A}\ }\textbf {\bibinfo {volume} {55}},\ \bibinfo {pages}
  {265201} (\bibinfo {year} {2022})},\ \Eprint
  {http://arxiv.org/abs/2107.10642} {arXiv:2107.10642 [hep-lat]} \BibitemShut
  {NoStop}%
\bibitem [{\citenamefont {Stetcu}\ \emph {et~al.}(2007)\citenamefont {Stetcu},
  \citenamefont {Barrett}, \citenamefont {van Kolck},\ and\ \citenamefont
  {Vary}}]{Stetcu:2007ms}%
  \BibitemOpen
  \bibfield  {author} {\bibinfo {author} {\bibfnamefont {I.}~\bibnamefont
  {Stetcu}}, \bibinfo {author} {\bibfnamefont {B.R.}\ \bibnamefont {Barrett}},
  \bibinfo {author} {\bibfnamefont {U.}~\bibnamefont {van Kolck}}, \ and\
  \bibinfo {author} {\bibfnamefont {J.P.}\ \bibnamefont {Vary}},\ }\bibfield
  {title} {\enquote {\bibinfo {title} {{Effective Theory for Trapped
  Few-Fermion Systems}},}\ }\href {\doibase 10.1103/PhysRevA.76.063613}
  {\bibfield  {journal} {\bibinfo  {journal} {Phys. Rev. A}\ }\textbf {\bibinfo
  {volume} {76}},\ \bibinfo {pages} {063613} (\bibinfo {year} {2007})},\
  \Eprint {http://arxiv.org/abs/0705.4335} {arXiv:0705.4335 [cond-mat.other]}
  \BibitemShut {NoStop}%
\bibitem [{\citenamefont {Stetcu}\ \emph {et~al.}(2010)\citenamefont {Stetcu},
  \citenamefont {Rotureau}, \citenamefont {Barrett},\ and\ \citenamefont {van
  Kolck}}]{Stetcu:2010xq}%
  \BibitemOpen
  \bibfield  {author} {\bibinfo {author} {\bibfnamefont {I.}~\bibnamefont
  {Stetcu}}, \bibinfo {author} {\bibfnamefont {J.}~\bibnamefont {Rotureau}},
  \bibinfo {author} {\bibfnamefont {B.R.}\ \bibnamefont {Barrett}}, \ and\
  \bibinfo {author} {\bibfnamefont {U.}~\bibnamefont {van Kolck}},\ }\bibfield
  {title} {\enquote {\bibinfo {title} {{An Effective field theory approach to
  two trapped particles}},}\ }\href {\doibase 10.1016/j.aop.2010.02.008}
  {\bibfield  {journal} {\bibinfo  {journal} {Annals Phys.}\ }\textbf {\bibinfo
  {volume} {325}},\ \bibinfo {pages} {1644--1666} (\bibinfo {year} {2010})},\
  \Eprint {http://arxiv.org/abs/1001.5071} {arXiv:1001.5071
  [cond-mat.quant-gas]} \BibitemShut {NoStop}%
\bibitem [{\citenamefont {Rotureau}\ \emph {et~al.}(2010)\citenamefont
  {Rotureau}, \citenamefont {Stetcu}, \citenamefont {Barrett}, \citenamefont
  {Birse},\ and\ \citenamefont {van Kolck}}]{Rotureau:2010uz}%
  \BibitemOpen
  \bibfield  {author} {\bibinfo {author} {\bibfnamefont {J.}~\bibnamefont
  {Rotureau}}, \bibinfo {author} {\bibfnamefont {I.}~\bibnamefont {Stetcu}},
  \bibinfo {author} {\bibfnamefont {B.R.}\ \bibnamefont {Barrett}}, \bibinfo
  {author} {\bibfnamefont {M.C.}\ \bibnamefont {Birse}}, \ and\ \bibinfo
  {author} {\bibfnamefont {U.}~\bibnamefont {van Kolck}},\ }\bibfield  {title}
  {\enquote {\bibinfo {title} {{Three and Four Harmonically Trapped Particles
  in an Effective Field Theory Framework}},}\ }\href {\doibase
  10.1103/PhysRevA.82.032711} {\bibfield  {journal} {\bibinfo  {journal} {Phys.
  Rev. A}\ }\textbf {\bibinfo {volume} {82}},\ \bibinfo {pages} {032711}
  (\bibinfo {year} {2010})},\ \Eprint {http://arxiv.org/abs/1006.3820}
  {arXiv:1006.3820 [cond-mat.quant-gas]} \BibitemShut {NoStop}%
\bibitem [{\citenamefont {Rotureau}\ \emph {et~al.}(2012)\citenamefont
  {Rotureau}, \citenamefont {Stetcu}, \citenamefont {Barrett},\ and\
  \citenamefont {van Kolck}}]{Rotureau:2011vf}%
  \BibitemOpen
  \bibfield  {author} {\bibinfo {author} {\bibfnamefont {J.}~\bibnamefont
  {Rotureau}}, \bibinfo {author} {\bibfnamefont {I.}~\bibnamefont {Stetcu}},
  \bibinfo {author} {\bibfnamefont {B.R.}\ \bibnamefont {Barrett}}, \ and\
  \bibinfo {author} {\bibfnamefont {U.}~\bibnamefont {van Kolck}},\ }\bibfield
  {title} {\enquote {\bibinfo {title} {{Two and Three Nucleons in a Trap and
  the Continuum Limit}},}\ }\href {\doibase 10.1103/PhysRevC.85.034003}
  {\bibfield  {journal} {\bibinfo  {journal} {Phys. Rev. C}\ }\textbf {\bibinfo
  {volume} {85}},\ \bibinfo {pages} {034003} (\bibinfo {year} {2012})},\
  \Eprint {http://arxiv.org/abs/1112.0267} {arXiv:1112.0267 [nucl-th]}
  \BibitemShut {NoStop}%
\bibitem [{\citenamefont {Luu}\ \emph {et~al.}(2010)\citenamefont {Luu},
  \citenamefont {Savage}, \citenamefont {Schwenk},\ and\ \citenamefont
  {Vary}}]{Luu:2010hw}%
  \BibitemOpen
  \bibfield  {author} {\bibinfo {author} {\bibfnamefont {Thomas}\ \bibnamefont
  {Luu}}, \bibinfo {author} {\bibfnamefont {Martin~J.}\ \bibnamefont {Savage}},
  \bibinfo {author} {\bibfnamefont {Achim}\ \bibnamefont {Schwenk}}, \ and\
  \bibinfo {author} {\bibfnamefont {James~P.}\ \bibnamefont {Vary}},\
  }\bibfield  {title} {\enquote {\bibinfo {title} {{Nucleon-Nucleon Scattering
  in a Harmonic Potential}},}\ }\href {\doibase 10.1103/PhysRevC.82.034003}
  {\bibfield  {journal} {\bibinfo  {journal} {Phys. Rev. C}\ }\textbf {\bibinfo
  {volume} {82}},\ \bibinfo {pages} {034003} (\bibinfo {year} {2010})},\
  \Eprint {http://arxiv.org/abs/1006.0427} {arXiv:1006.0427 [nucl-th]}
  \BibitemShut {NoStop}%
\bibitem [{\citenamefont {Yang}(2016)}]{Yang:2016brl}%
  \BibitemOpen
  \bibfield  {author} {\bibinfo {author} {\bibfnamefont {C.-J.}\ \bibnamefont
  {Yang}},\ }\bibfield  {title} {\enquote {\bibinfo {title} {{Chiral potential
  renormalized in harmonic-oscillator space}},}\ }\href {\doibase
  10.1103/PhysRevC.94.064004} {\bibfield  {journal} {\bibinfo  {journal} {Phys.
  Rev. C}\ }\textbf {\bibinfo {volume} {94}},\ \bibinfo {pages} {064004}
  (\bibinfo {year} {2016})},\ \Eprint {http://arxiv.org/abs/1610.01350}
  {arXiv:1610.01350 [nucl-th]} \BibitemShut {NoStop}%
\bibitem [{\citenamefont {Johnson}\ \emph {et~al.}(2019)\citenamefont {Johnson}
  \emph {et~al.}}]{Johnson:2019sps}%
  \BibitemOpen
  \bibfield  {author} {\bibinfo {author} {\bibfnamefont {Calvin~W.}\
  \bibnamefont {Johnson}} \emph {et~al.},\ }\bibfield  {title} {\enquote
  {\bibinfo {title} {{From bound states to the continuum}},}\ }in\ \href@noop
  {} {\emph {\bibinfo {booktitle} {{From Bound States to the Continuum}:
  {Connecting bound state calculations with scattering and reaction theory}}}}\
  (\bibinfo {year} {2019})\ \Eprint {http://arxiv.org/abs/1912.00451}
  {arXiv:1912.00451 [nucl-th]} \BibitemShut {NoStop}%
\bibitem [{\citenamefont {Zhang}(2020)}]{Zhang:2019cai}%
  \BibitemOpen
  \bibfield  {author} {\bibinfo {author} {\bibfnamefont {Xilin}\ \bibnamefont
  {Zhang}},\ }\bibfield  {title} {\enquote {\bibinfo {title} {{Extracting
  free-space observables from trapped interacting clusters}},}\ }\href
  {\doibase 10.1103/PhysRevC.101.051602} {\bibfield  {journal} {\bibinfo
  {journal} {Phys. Rev. C}\ }\textbf {\bibinfo {volume} {101}},\ \bibinfo
  {pages} {051602} (\bibinfo {year} {2020})},\ \Eprint
  {http://arxiv.org/abs/1905.05275} {arXiv:1905.05275 [nucl-th]} \BibitemShut
  {NoStop}%
\bibitem [{\citenamefont {Zhang}\ \emph {et~al.}(2020)\citenamefont {Zhang},
  \citenamefont {Stroberg}, \citenamefont {Navr\'atil}, \citenamefont {Gwak},
  \citenamefont {Melendez}, \citenamefont {Furnstahl},\ and\ \citenamefont
  {Holt}}]{Zhang:2020rhz}%
  \BibitemOpen
  \bibfield  {author} {\bibinfo {author} {\bibfnamefont {Xilin}\ \bibnamefont
  {Zhang}}, \bibinfo {author} {\bibfnamefont {S.R.}\ \bibnamefont {Stroberg}},
  \bibinfo {author} {\bibfnamefont {P.}~\bibnamefont {Navr\'atil}}, \bibinfo
  {author} {\bibfnamefont {Chan}\ \bibnamefont {Gwak}}, \bibinfo {author}
  {\bibfnamefont {J.A.}\ \bibnamefont {Melendez}}, \bibinfo {author}
  {\bibfnamefont {R.J.}\ \bibnamefont {Furnstahl}}, \ and\ \bibinfo {author}
  {\bibfnamefont {J.D.}\ \bibnamefont {Holt}},\ }\bibfield  {title} {\enquote
  {\bibinfo {title} {{Ab initio calculations of low-energy nuclear scattering
  using a generalized L\"uscher method}},}\ }\href {\doibase
  10.1103/PhysRevLett.125.112503} {\bibfield  {journal} {\bibinfo  {journal}
  {Phys. Rev. Lett.}\ }\textbf {\bibinfo {volume} {125}},\ \bibinfo {pages}
  {112503} (\bibinfo {year} {2020})},\ \Eprint
  {http://arxiv.org/abs/2004.13575} {arXiv:2004.13575 [nucl-th]} \BibitemShut
  {NoStop}%
\bibitem [{\citenamefont {Wang}\ \emph {et~al.}(2024)\citenamefont {Wang},
  \citenamefont {Du}, \citenamefont {Zuo},\ and\ \citenamefont
  {Vary}}]{PhysRevC.109.064623}%
  \BibitemOpen
  \bibfield  {author} {\bibinfo {author} {\bibfnamefont {Peiyan}\ \bibnamefont
  {Wang}}, \bibinfo {author} {\bibfnamefont {Weijie}\ \bibnamefont {Du}},
  \bibinfo {author} {\bibfnamefont {Wei}\ \bibnamefont {Zuo}}, \ and\ \bibinfo
  {author} {\bibfnamefont {James~P.}\ \bibnamefont {Vary}},\ }\bibfield
  {title} {\enquote {\bibinfo {title} {{Nuclear scattering via quantum
  computing}},}\ }\href {\doibase 10.1103/PhysRevC.109.064623} {\bibfield
  {journal} {\bibinfo  {journal} {Phys. Rev. C}\ }\textbf {\bibinfo {volume}
  {109}},\ \bibinfo {pages} {064623} (\bibinfo {year} {2024})},\ \Eprint
  {http://arxiv.org/abs/2401.17138} {arXiv:2401.17138 [nucl-th]} \BibitemShut
  {NoStop}%
\bibitem [{\citenamefont {Turro}\ \emph {et~al.}(2024)\citenamefont {Turro},
  \citenamefont {Wendt}, \citenamefont {Quaglioni}, \citenamefont {Pederiva},\
  and\ \citenamefont {Roggero}}]{PhysRevC.110.054604}%
  \BibitemOpen
  \bibfield  {author} {\bibinfo {author} {\bibfnamefont {Francesco}\
  \bibnamefont {Turro}}, \bibinfo {author} {\bibfnamefont {Kyle~A.}\
  \bibnamefont {Wendt}}, \bibinfo {author} {\bibfnamefont {Sofia}\ \bibnamefont
  {Quaglioni}}, \bibinfo {author} {\bibfnamefont {Francesco}\ \bibnamefont
  {Pederiva}}, \ and\ \bibinfo {author} {\bibfnamefont {Alessandro}\
  \bibnamefont {Roggero}},\ }\bibfield  {title} {\enquote {\bibinfo {title}
  {{Evaluation of phase shifts for nonrelativistic elastic scattering using
  quantum computers}},}\ }\href {\doibase 10.1103/PhysRevC.110.054604}
  {\bibfield  {journal} {\bibinfo  {journal} {Phys. Rev. C}\ }\textbf {\bibinfo
  {volume} {110}},\ \bibinfo {pages} {054604} (\bibinfo {year} {2024})},\
  \Eprint {http://arxiv.org/abs/2407.04155} {arXiv:2407.04155 [quant-ph]}
  \BibitemShut {NoStop}%
\bibitem [{\citenamefont {Gustafson}\ \emph {et~al.}(2021)\citenamefont
  {Gustafson}, \citenamefont {Zhu}, \citenamefont {Dreher}, \citenamefont
  {Linke},\ and\ \citenamefont {Meurice}}]{PhysRevD.104.054507}%
  \BibitemOpen
  \bibfield  {author} {\bibinfo {author} {\bibfnamefont {Erik}\ \bibnamefont
  {Gustafson}}, \bibinfo {author} {\bibfnamefont {Yingyue}\ \bibnamefont
  {Zhu}}, \bibinfo {author} {\bibfnamefont {Patrick}\ \bibnamefont {Dreher}},
  \bibinfo {author} {\bibfnamefont {Norbert~M.}\ \bibnamefont {Linke}}, \ and\
  \bibinfo {author} {\bibfnamefont {Yannick}\ \bibnamefont {Meurice}},\
  }\bibfield  {title} {\enquote {\bibinfo {title} {Real-time quantum
  calculations of phase shifts using wave packet time delays},}\ }\href
  {\doibase 10.1103/PhysRevD.104.054507} {\bibfield  {journal} {\bibinfo
  {journal} {Phys. Rev. D}\ }\textbf {\bibinfo {volume} {104}},\ \bibinfo
  {pages} {054507} (\bibinfo {year} {2021})}\BibitemShut {NoStop}%
\bibitem [{\citenamefont {Mussardo}\ \emph {et~al.}(2024)\citenamefont
  {Mussardo}, \citenamefont {Stampiggi},\ and\ \citenamefont
  {Trombettoni}}]{Mussardo2024}%
  \BibitemOpen
  \bibfield  {author} {\bibinfo {author} {\bibfnamefont {Giuseppe}\
  \bibnamefont {Mussardo}}, \bibinfo {author} {\bibfnamefont {Andrea}\
  \bibnamefont {Stampiggi}}, \ and\ \bibinfo {author} {\bibfnamefont {Andrea}\
  \bibnamefont {Trombettoni}},\ }\bibfield  {title} {\enquote {\bibinfo {title}
  {Reflection and transmission amplitudes in a digital quantum simulation},}\
  }\href {\doibase 10.1140/epjqt/s40507-024-00277-3} {\bibfield  {journal}
  {\bibinfo  {journal} {EPJ Quantum Technology}\ }\textbf {\bibinfo {volume}
  {11}},\ \bibinfo {pages} {65} (\bibinfo {year} {2024})}\BibitemShut {NoStop}%
\bibitem [{\citenamefont {Guo}\ and\ \citenamefont
  {Gasparian}(2023)}]{Guo:2023ecc}%
  \BibitemOpen
  \bibfield  {author} {\bibinfo {author} {\bibfnamefont {Peng}\ \bibnamefont
  {Guo}}\ and\ \bibinfo {author} {\bibfnamefont {Vladimir}\ \bibnamefont
  {Gasparian}},\ }\bibfield  {title} {\enquote {\bibinfo {title} {{Toward
  extracting the scattering phase shift from integrated correlation
  functions}},}\ }\href {\doibase 10.1103/PhysRevD.108.074504} {\bibfield
  {journal} {\bibinfo  {journal} {Phys. Rev. D}\ }\textbf {\bibinfo {volume}
  {108}},\ \bibinfo {pages} {074504} (\bibinfo {year} {2023})},\ \Eprint
  {http://arxiv.org/abs/2307.12951} {arXiv:2307.12951 [hep-lat]} \BibitemShut
  {NoStop}%
\bibitem [{\citenamefont {Guo}(2024)}]{Guo:2024zal}%
  \BibitemOpen
  \bibfield  {author} {\bibinfo {author} {\bibfnamefont {Peng}\ \bibnamefont
  {Guo}},\ }\bibfield  {title} {\enquote {\bibinfo {title} {{Toward extracting
  the scattering phase shift from integrated correlation functions. II. A
  relativistic lattice field theory model}},}\ }\href {\doibase
  10.1103/PhysRevD.110.014504} {\bibfield  {journal} {\bibinfo  {journal}
  {Phys. Rev. D}\ }\textbf {\bibinfo {volume} {110}},\ \bibinfo {pages}
  {014504} (\bibinfo {year} {2024})},\ \Eprint
  {http://arxiv.org/abs/2402.15628} {arXiv:2402.15628 [hep-lat]} \BibitemShut
  {NoStop}%
\bibitem [{\citenamefont {Guo}\ and\ \citenamefont {Lee}(2025)}]{Guo:2024pvt}%
  \BibitemOpen
  \bibfield  {author} {\bibinfo {author} {\bibfnamefont {Peng}\ \bibnamefont
  {Guo}}\ and\ \bibinfo {author} {\bibfnamefont {Frank~X.}\ \bibnamefont
  {Lee}},\ }\bibfield  {title} {\enquote {\bibinfo {title} {{Toward extracting
  scattering phase shift from integrated correlation functions. III. Coupled
  channels}},}\ }\href {\doibase 10.1103/PhysRevD.111.054506} {\bibfield
  {journal} {\bibinfo  {journal} {Phys. Rev. D}\ }\textbf {\bibinfo {volume}
  {111}},\ \bibinfo {pages} {054506} (\bibinfo {year} {2025})},\ \Eprint
  {http://arxiv.org/abs/2412.00812} {arXiv:2412.00812 [hep-lat]} \BibitemShut
  {NoStop}%
\bibitem [{\citenamefont {Ortiz}\ \emph {et~al.}(2001)\citenamefont {Ortiz},
  \citenamefont {Gubernatis}, \citenamefont {Knill},\ and\ \citenamefont
  {Laflamme}}]{PhysRevA.64.022319}%
  \BibitemOpen
  \bibfield  {author} {\bibinfo {author} {\bibfnamefont {G.}~\bibnamefont
  {Ortiz}}, \bibinfo {author} {\bibfnamefont {J.~E.}\ \bibnamefont
  {Gubernatis}}, \bibinfo {author} {\bibfnamefont {E.}~\bibnamefont {Knill}}, \
  and\ \bibinfo {author} {\bibfnamefont {R.}~\bibnamefont {Laflamme}},\
  }\bibfield  {title} {\enquote {\bibinfo {title} {Quantum algorithms for
  fermionic simulations},}\ }\href {\doibase 10.1103/PhysRevA.64.022319}
  {\bibfield  {journal} {\bibinfo  {journal} {Phys. Rev. A}\ }\textbf {\bibinfo
  {volume} {64}},\ \bibinfo {pages} {022319} (\bibinfo {year}
  {2001})}\BibitemShut {NoStop}%
\bibitem [{\citenamefont {Somma}\ \emph {et~al.}(2002)\citenamefont {Somma},
  \citenamefont {Ortiz}, \citenamefont {Gubernatis}, \citenamefont {Knill},\
  and\ \citenamefont {Laflamme}}]{PhysRevA.65.042323}%
  \BibitemOpen
  \bibfield  {author} {\bibinfo {author} {\bibfnamefont {R.}~\bibnamefont
  {Somma}}, \bibinfo {author} {\bibfnamefont {G.}~\bibnamefont {Ortiz}},
  \bibinfo {author} {\bibfnamefont {J.~E.}\ \bibnamefont {Gubernatis}},
  \bibinfo {author} {\bibfnamefont {E.}~\bibnamefont {Knill}}, \ and\ \bibinfo
  {author} {\bibfnamefont {R.}~\bibnamefont {Laflamme}},\ }\bibfield  {title}
  {\enquote {\bibinfo {title} {Simulating physical phenomena by quantum
  networks},}\ }\href {\doibase 10.1103/PhysRevA.65.042323} {\bibfield
  {journal} {\bibinfo  {journal} {Phys. Rev. A}\ }\textbf {\bibinfo {volume}
  {65}},\ \bibinfo {pages} {042323} (\bibinfo {year} {2002})}\BibitemShut
  {NoStop}%
\bibitem [{\citenamefont {Jordan}\ and\ \citenamefont
  {Wigner}(1928)}]{JordanWigner}%
  \BibitemOpen
  \bibfield  {author} {\bibinfo {author} {\bibfnamefont {P.}~\bibnamefont
  {Jordan}}\ and\ \bibinfo {author} {\bibfnamefont {E.}~\bibnamefont
  {Wigner}},\ }\bibfield  {title} {\enquote {\bibinfo {title} {{\"U}ber das
  paulische {\"a}quivalenzverbot},}\ }\href {\doibase 10.1007/BF01331938}
  {\bibfield  {journal} {\bibinfo  {journal} {Zeitschrift f{\"u}r Physik}\
  }\textbf {\bibinfo {volume} {47}},\ \bibinfo {pages} {631--651} (\bibinfo
  {year} {1928})}\BibitemShut {NoStop}%
\bibitem [{\citenamefont {Tacchino}\ \emph {et~al.}(2020)\citenamefont
  {Tacchino}, \citenamefont {Chiesa}, \citenamefont {Carretta},\ and\
  \citenamefont {Gerace}}]{https://doi.org/10.1002/qute.201900052}%
  \BibitemOpen
  \bibfield  {author} {\bibinfo {author} {\bibfnamefont {Francesco}\
  \bibnamefont {Tacchino}}, \bibinfo {author} {\bibfnamefont {Alessandro}\
  \bibnamefont {Chiesa}}, \bibinfo {author} {\bibfnamefont {Stefano}\
  \bibnamefont {Carretta}}, \ and\ \bibinfo {author} {\bibfnamefont {Dario}\
  \bibnamefont {Gerace}},\ }\bibfield  {title} {\enquote {\bibinfo {title}
  {Quantum computers as universal quantum simulators: State-of-the-art and
  perspectives},}\ }\href {\doibase https://doi.org/10.1002/qute.201900052}
  {\bibfield  {journal} {\bibinfo  {journal} {Advanced Quantum Technologies}\
  }\textbf {\bibinfo {volume} {3}},\ \bibinfo {pages} {1900052} (\bibinfo
  {year} {2020})}\BibitemShut {NoStop}%
\bibitem [{\citenamefont {Trotter}(1959)}]{TrotterH.F.1959Otpo}%
  \BibitemOpen
  \bibfield  {author} {\bibinfo {author} {\bibfnamefont {H.~F.}\ \bibnamefont
  {Trotter}},\ }\bibfield  {title} {\enquote {\bibinfo {title} {On the product
  of semi-groups of operators},}\ }\href {http://www.jstor.org/stable/2033649}
  {\bibfield  {journal} {\bibinfo  {journal} {Proceedings of the American
  Mathematical Society}\ }\textbf {\bibinfo {volume} {10}},\ \bibinfo {pages}
  {545--551} (\bibinfo {year} {1959})}\BibitemShut {NoStop}%
\bibitem [{\citenamefont {Hatano}\ and\ \citenamefont
  {Suzuki}(2005)}]{Hatano:2005gh}%
  \BibitemOpen
  \bibfield  {author} {\bibinfo {author} {\bibfnamefont {Naomichi}\
  \bibnamefont {Hatano}}\ and\ \bibinfo {author} {\bibfnamefont {Masuo}\
  \bibnamefont {Suzuki}},\ }\bibfield  {title} {\enquote {\bibinfo {title}
  {{Finding Exponential Product Formulas of Higher Orders}},}\ }\href {\doibase
  10.1007/11526216_2} {\bibfield  {journal} {\bibinfo  {journal} {Lect. Notes
  Phys.}\ }\textbf {\bibinfo {volume} {679}},\ \bibinfo {pages} {37} (\bibinfo
  {year} {2005})},\ \Eprint {http://arxiv.org/abs/math-ph/0506007}
  {arXiv:math-ph/0506007} \BibitemShut {NoStop}%
\bibitem [{\citenamefont {Guo}\ \emph {et~al.}(2023)\citenamefont {Guo},
  \citenamefont {Gasparian}, \citenamefont {J\'odar},\ and\ \citenamefont
  {Wisehart}}]{Guo:2022jyk}%
  \BibitemOpen
  \bibfield  {author} {\bibinfo {author} {\bibfnamefont {Peng}\ \bibnamefont
  {Guo}}, \bibinfo {author} {\bibfnamefont {Vladimir}\ \bibnamefont
  {Gasparian}}, \bibinfo {author} {\bibfnamefont {Esther}\ \bibnamefont
  {J\'odar}}, \ and\ \bibinfo {author} {\bibfnamefont {Christopher}\
  \bibnamefont {Wisehart}},\ }\bibfield  {title} {\enquote {\bibinfo {title}
  {{Tunneling time in PT-symmetric systems}},}\ }\href {\doibase
  10.1103/PhysRevA.107.032210} {\bibfield  {journal} {\bibinfo  {journal}
  {Phys. Rev. A}\ }\textbf {\bibinfo {volume} {107}},\ \bibinfo {pages}
  {032210} (\bibinfo {year} {2023})},\ \Eprint
  {http://arxiv.org/abs/2208.13543} {arXiv:2208.13543 [cond-mat.other]}
  \BibitemShut {NoStop}%
\end{thebibliography}%

\end{document}